\documentclass[3p,times,procedia]{elsarticle}
\usepackage{nupha_ecrc}

\volume{00}

\firstpage{1}

\journalname{Nuclear Physics A}

\runauth{}

\jid{nupha}

\jnltitlelogo{Nuclear Physics A}

\usepackage{amssymb}

\usepackage[figuresright]{rotating}

\usepackage{amssymb,mathtools}
\usepackage{amsthm}
\usepackage{bm}
\usepackage{graphicx}
\usepackage{epstopdf}

\usepackage[figuresright]{rotating}

\newcommand{\beq}{\begin{eqnarray}}
\newcommand{\eeq}{\end{eqnarray}}

\newcommand{\bal}{\begin{align}}
\newcommand{\eal}{\begin{align}}

\def\Tjet{\Theta_\text{jet} }

\newcommand{\rmd}{{\rm d}}

\def\q{{\bm q}}
\def\0{{\boldsymbol 0}}

\def\k{{\boldsymbol k}}

\def\x{{\boldsymbol x}}

\newcommand{\del}{\partial}

\begin{document}

\begin{frontmatter}



\dochead{}

\title{Theoretical Developments in QCD Jet Energy Loss}


\author{Yacine Mehtar-Tani}

\address{Institute for Nuclear Theory, University of Washington, \\Seattle, WA 98195-1550, USA}

\begin{abstract}
We review the recent developments in the theory of jet-quenching. First, we analyze the coherent vacuum cascade and incoherent medium-induced cascade separately. We then discuss the interplay between the two kinds of cascade and the resulting partial decoherence of the inner jet structure. Finally, we report on recent calculations of higher-order corrections. In particular, the dominant radiative corrections to jet observables that yield the renormalization of the jet-quenching parameter are addressed. 
 \end{abstract}


\end{frontmatter}
\vspace{0.5cm}
\begin{flushleft}
\end{flushleft}
\begin{flushright}
INT-PUB-16-002
\end{flushright}


\section{Introduction}
\label{sec:intro}
The most important discovery of heavy ion programs at RHIC and the LHC is undoubtedly that a deconfined matter behaving as a perfect liquid is created in ultra-relativistic heavy ion collisions (see Ref.~\cite{d'Enterria:2009am} for an experimental review).  The formation of a deconfined QCD matter in high energy hadronic collisions was suggested long ago. In a seminal work \cite{Bjorken:1982tu}, Bjorken argued that such new state of matter could be signaled by the suppression of high-pt particle spectra compared to proton-proton collisions (after a proper rescaling by the number of binary collisions).  This phenomenon was successfully observed in ultra-relativistic heavy-ion collisions at RHIC and LHC and is commonly referred to as ``Jet-quenching". 

Jet-quenching is attributed to the energy loss of energetic partons while propagating through the QCD medium produced in the collision. Two  perturbative mechanisms were put forward to quantitatively account for the observed suppression: i) Elastic parton energy loss, in which energy is lost to the QCD medium via elastic rescatterings of the hard partons off the medium color charges \cite{Braaten:1991we,Djordjevic:2006tw}. ii) Inelastic parton energy loss \cite{Gyulassy:1993hr,Baier:1996kr,Zakharov:1996fv,Gyulassy:2000er,Wiedemann:2000za,Guo:2000nz,Baier:2001yt,Arnold:2002ja}, where energy is lost inelastically by medium-induced soft gluon radiation. The average elastic energy loss increases linearly as a function of the medium length $L$, while radiative energy loss increases quadratically, which makes the latter the dominant source of energy loss for large media. \footnote{In the case of heavy-quark production inelastic processes are suppressed due to the dead-cone effect and elastic processes are no longer negligible.}

Until recently, Jet-quenching phenomenology was limited to the study of inclusive hadron spectra at moderate pt's ($5-15$ GeV). The energies reached by the LHC, allowed for the first time the production of jets in heavy-ion collisions in large numbers \cite{Aad:2010bu}. Jets are collimated beams of energetic particles and as such encode more information than single inclusive particle spectra, providing a more refined diagnostic tool for probing the QGP. Although the theory of parton energy loss has been successful in describing single particle spectra, it was not designed to describe in-medium parton showers, and hence jets, mainly because it is based on leading order calculations, namely, the single gluon radiation spectrum.  In the absence of a consistent theory of jets in a medium, phenomenological studies of jet observables relied to a large extend on modeling of medium modified parton showers. As a result, there exists a multitude of Monte Carlo even generators, based on different assumptions as to how to extend the leading order medium-induced gluon radiation spectrum to multiple parton branchings \cite{Zapp:2011ya}. 

In this contribution, we report on recent progress toward understanding the space-time structure of parton showers in the presence of QCD medium (see Refs.~\cite{Mehtar-Tani:2013pia,Blaizot:2015lma} for recent reviews). We shall leave aside practical questions concerning, for instance, medium evolution, geometry, kinematics, particle species dependence, Monte Carlo implementation, etc, and focus on conceptual issues regarding the structure of medium-modified parton shower. 

\section{Parton cascades of two kinds}
\label{sec:twokinds}
In vacuum, a jet is described by a virtuality ordered and collimated parton shower. In the presence of a dense medium, another type of parton shower forms induced by  multiple  interactions with medium. This medium-induced parton cascade is of a fundamentally different nature than the vacuum shower as it evolves in real time and develops at large angles.  
\subsection{Vacuum cascade}
\label{sec:vacuum}
\begin{figure}[htbp]\label{fig:vacuum}
\centering
\includegraphics[width=7cm]{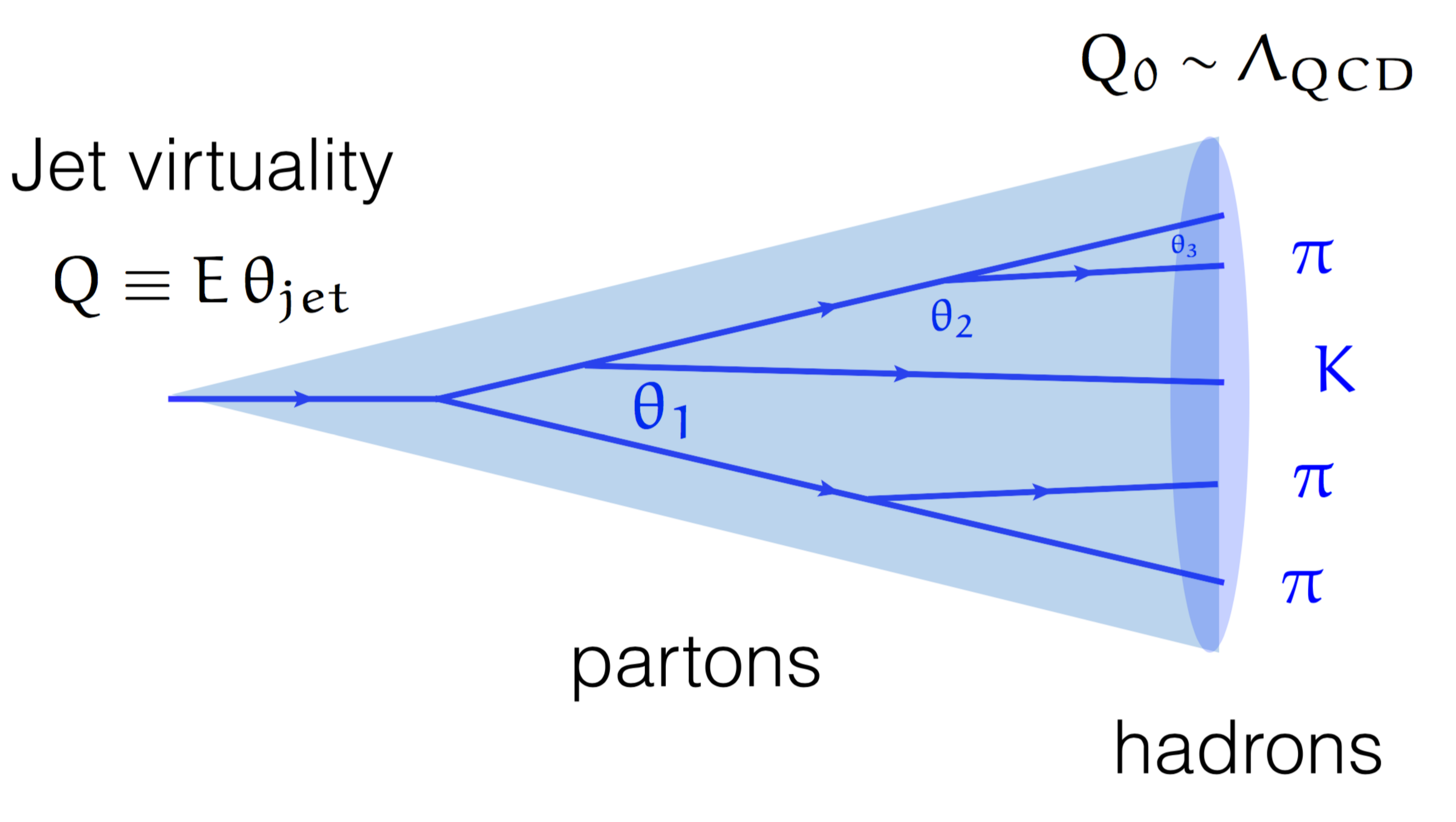} 
\caption{Illustration of the vacuum casc`ade:  three successive branchings, ordered in angle, are depicted: $ \theta_3 \ll \theta_2 \ll \theta_1 \ll \Tjet$. Virtuality decreases from $Q$ down to a non-perturbative scale $Q_0$ where hadronization takes place. }\end{figure}

A jet originates from a high energy parton created in a hard process. Because of its high virtuality the energetic parent parton tends to branch with a probability that is proportional to the coupling constant and enhanced by collinear and soft logarithmic singularities ($\theta \to 0$, $\omega\to 0 $), 
\beq\label{rad-prob}
\rmd P\sim \bar\alpha \frac{\rmd \omega }{\omega}\frac{\rmd \theta}{\theta}, 
\eeq 
with $\bar\alpha \equiv \alpha_s N_c/\pi$ for gluon branching. This elementary process captures two characteristic features of QCD jets, namely that multiple parton branching is highly probable and collimated along the original parton momentum direction. The dominant logarithmic contribution to inclusive jet observables, such as the fragmentation function, is given by strongly ordered successive branchings in virtuality from high $Q^2 \equiv E \Tjet $ (where $E$ is the jet energy and $\Tjet$ is the jet opening angle, often referred to as $p_t$ and $R$, respectively, in the literature) to low virtuality, down to a non-perturbative scale $Q_0\sim \Lambda_\text{QCD}$, where hadronization takes place converting partons to hadrons. This ordering results from a strong ordering in formation time along the parton cascade \cite{Dokshitzer:1991wu}. However, due to interference effects, the parton shower does not reduce to a plain iteration of the 1 to 2 parton branching probability ordered in virtuality or transverse momentum. The jet is a color coherent system of partons,  and as a consequence, large angle soft gluon radiation, characterized by a transverse wave length $\lambda_\perp\sim 1/k_\perp$  larger than the transverse size of the  emitting system, does not ``resolve" the individual radiating color charges. The resulting radiation intensity is suppressed compared to incoherent radiation off individual charges. In turn, it is proportional to the total color charge that is that of the original parton. Color coherence of the parton shower  is a consequence of color charge conservation and can be accounted for by imposing a strict angular ordering of successive parton branchings in place of the less restrictive virtuality ordering prescription \cite{Bassetto:1982ma,Bassetto:1984ik} (see Fig.~\ref{fig:vacuum}). Therefore, the QCD evolution equation that describes the evolution of the fragmentation function, the Modified-Leading-Log-Approximation (MLLA) equation differs from the DGLAP equation in the choice of the jet angle as the correct evolution variable instead of the jet virtuality \cite{Dokshitzer:1991wu}. 

To summarize, jets in vacuum are characterized by a strongly collimated parton shower and a suppression of large angle soft gluon radiation. These important features are to be contrasted with the medium-induced shower that we shall discuss in the following section. 

\subsection{Medium-induced cascade}
\label{sec:medium}
A high energy parton passing through a hot QCD medium undergoes multiple scatterings.  In the weak coupling regime the in-medium correlation length given by the inverse Debye mass, $1/m_D\sim 1/gT$ is much smaller than the typical in-medium mean-free-path, $\ell_\text{mfp}\sim 1/g^2T$, where $g\ll 1$ is the coupling constant and $T \gg \Lambda_\text{QCD}$ the temperature of the QGP. This allows us to treat multiple scatterings as effectively independent. Hence, the coupling of energetic partons with the medium can be characterized by local transport coefficients, the most significant of which are the so-called jet-quenching parameter, $\hat q\equiv \rmd \langle k_\perp^2\rangle /\rmd t$ \cite{Baier:1996kr} and $\hat e \equiv \rmd  \langle E\rangle/ \rmd t$ \cite{Majumder:2008zg},  that are related to transverse momentum broadening and collisional energy loss, respectively.  

\begin{figure}[htbp]\label{fig:medium}
\centering
\includegraphics[width=8cm]{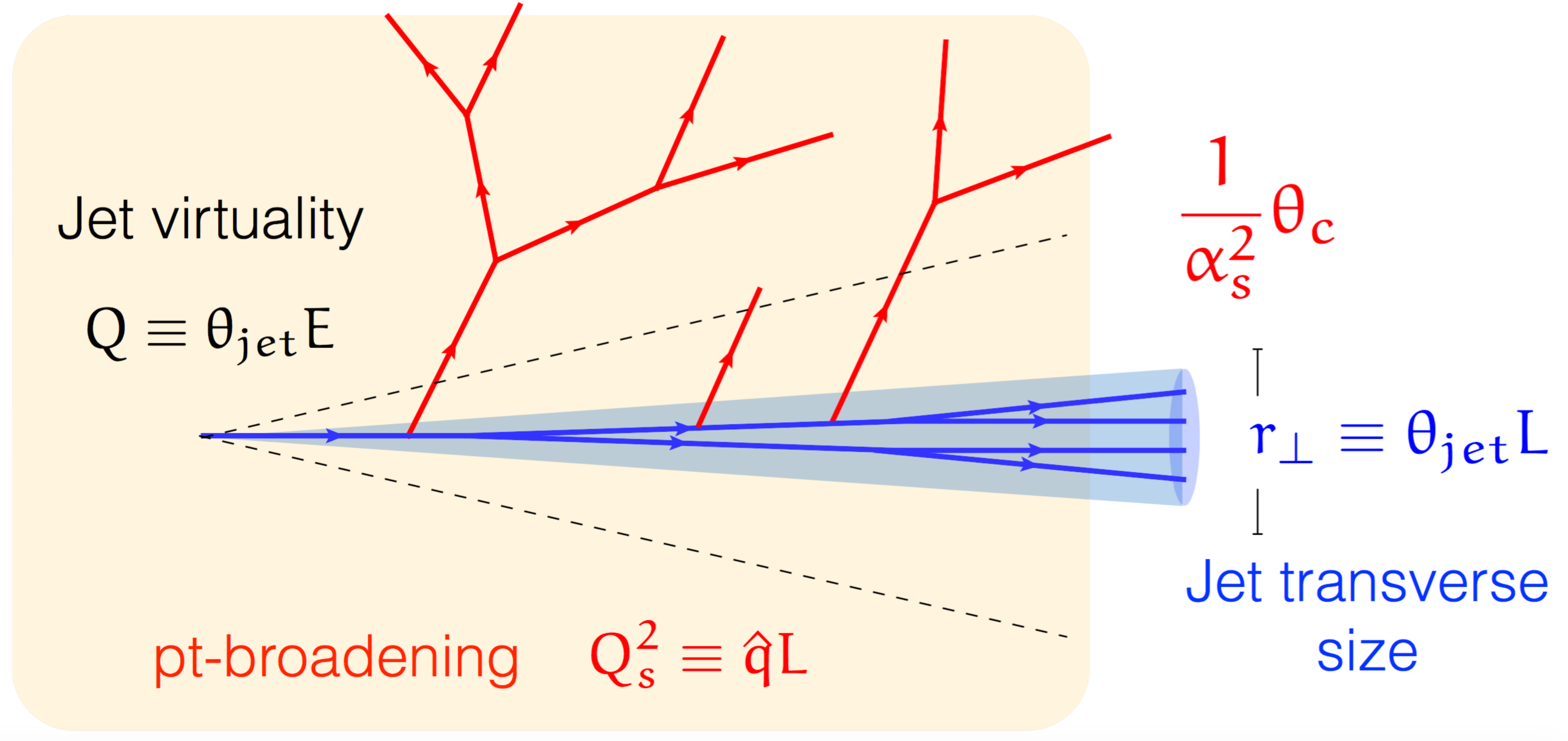} 

\caption{An illustration of large angle in-medium turbulent cascade (red) and collimated vacuum cascade (blue). In the limit $r_\perp \ll Q_s^{-1}$, or equivalently when $\Tjet \ll \theta_c$, the medium-induced shower is induced by the total charge of the jet core and hence, is not sensitive to the internal structure of the jet which evolves coherently as in vacuum apart from an overall energy reduction due to medium induced radiation (see text for details).} \end{figure}

Multiple interactions of the hard parton with the medium constituents can trigger gluon radiation coherently over a time $t_f\sim \omega/k^2_\perp\sim \sqrt{\omega/\hat q}$, where $\omega$ is the frequency of the radiated gluon. For $\ell_\text{mfp}\ll t_f(\omega) \ll L$, the radiation probability reads (cf. Ref.~\cite{Peigne:2008wu} for a review)
\beq\label{med-ind-prob}
\rmd P \sim \bar\alpha \frac{\rmd \omega }{\omega}\,  \frac{\rmd t}{ t_f(\omega)},
\eeq
where $t$ runs up to the length of the medium $L$. Note that hard gluon radiation, corresponding to large formation times, are suppressed compared to incoherent radiation whose rate is $\omega \rmd P/\rmd \omega \rmd t\sim \alpha_s/\ell_\text{mfp}$. This is the QCD analog of the Landau-Pomeranchuk-Migdal (LPM) effect. The maximum suppression is achieved when $t_f(\omega)\sim L$. Hence, gluons with frequencies larger than the characteristic frequency $\omega_c\equiv \hat q L^2$ are strongly suppressed. This upper bound in energies coincides with a lower radiation angle for medium-induced radiation: $\theta_c \equiv Q_s  /\omega_c \sim 1/\sqrt{\hat q L^3}$ where $Q_s^2 \sim \hat q t_f(\omega_c)\sim \hat q L$ is the typical  transverse momentum broadening that the radiated gluon acquires in the medium. In Eq.~(\ref{med-ind-prob}), we observe that multiple branching become highly probable when the characteristic time scale 
\beq
t_\ast(\omega)\, \equiv\,  \frac{1}{\bar\alpha}\sqrt{\frac{\omega}{\hat q }}\, \equiv \, \frac{1}{\bar\alpha}t_f(\omega) \,   \ll\,  L.
\eeq 
$t_\ast(\omega)$ corresponds to the time it takes for a parton to branch in the medium, with $\omega$ the softest frequency involved in the branching process. This yields a characteristic frequency $\omega_s\sim \bar\alpha^2 \omega_c$ and a minimum angle \cite{Blaizot:2014ula,Kurkela:2014tla} 
\beq
\theta_s \equiv \frac{1}{\bar\alpha^2} \theta_c \gg \theta_c,
\eeq
for the multiple branching regime (cf. Fig.~\ref{fig:medium}). In this regime, the time between two branchings $t_\ast(\omega)$ is parametrically larger than the time it takes for a branching to happen, $t_f(\omega)$, and because of in-medium rapid color randomization of the off-spring gluons, multiple branchings can be treated as effectively independent \cite{Blaizot:2012fh,Apolinario:2014csa}. This separation of time scales allows for a probabilistic description of in-medium parton shower characterized by the real-time evolution variable $t\sim L$ and a quasi-local branching probability \cite{Blaizot:2013vha}. 

An exemple of observables that can be computed in this framework is the inclusive single gluon distribution: $D(\omega) \equiv \omega \rmd N/ \rmd \omega $, which obeys the following rate equation (in the pure gluonic case) \cite{Baier:2000sb,Jeon:2003gi,Blaizot:2013vha,Blaizot:2013hx}, 
\beq\label{rate-eq}
\frac{\del }{\del t} D(\omega) = \int_0^1 \rmd z\, {\cal K}(z) \left[ \frac{D(\omega/z)}{t_\ast(\omega/z)}-z\frac{D(\omega)}{t_\ast(\omega)}\right],
\eeq
where the kernel $\cal K$ generalizes the radiation probability (\ref{med-ind-prob}) by encompassing energy conservation, ${\cal K}(z) \sim  [1-z+z^2]^{5/2}/[z(1-z)]^{3/2}$. The first term in Eq.~(\ref{rate-eq}) corresponds to the rate for gaining a gluon with frequency $\omega$ from the decay of a parent gluon with frequency $\omega/z$ whereas the second term corresponds to its loss rate by decaying into two gluons. The kernel $\cal K$ was first derived in the infinite medium limit and was generalized recently including finite size effects \cite{Apolinario:2014csa}. The gluon cascade described by  Eq.~(\ref{rate-eq}) has remarkable features. It admits a scaling solution of the form $D(\omega)\sim \tau/\sqrt{\omega} \exp[-\pi\tau^2]$ for $\omega \ll E$, with $\tau \equiv L/t_\ast(E)$\footnote{This exact solution was derived for the simplified kernel ${\cal K}(z) \equiv 1/[z(1-z)]^{3/2}$.} corresponding to a finite flux of energy  $F \simeq \pi \omega_s $, at the origine $\omega=0$. This transport of energy from energy containing partons to arbitrarily low frequencies with a constant flux is reminiscent of wave turbulence \cite{Blaizot:2013hx,Fister:2014zxa,Blaizot:2015jea}.  In fact, energy must dissipate when $\omega\sim T \ll E$, the temperature of the plasma \cite{Iancu:2015uja}. The angular distribution of the turbulent cascade in the multiple branching and scattering regime was also investigated, and was shown to be close to a Gaussian with a variance given by the characteristic angle $\theta^2_\ast(\omega)\equiv (\hat q/\omega^3)^{1/2}/4\bar\alpha > \theta^2_s$ \cite{Blaizot:2014ula,Blaizot:2014rla}.  
The turbulent energy transport mechanism from the forward partons to soft gluons at large angles provides a possible explanation for the observed missing energy in asymmetric dijet events \cite{Blaizot:2014ula}, where energy was recovered only at large angles and in soft particles by the CMS collaboration \cite{CMS:2014uca}.

\subsection{A simple pQCD limit: coherent jets}
\label{sec:coherent}

In the previous sections, we have discussed vacuum and medium-induced showers separately. A complete theory of jets in a QCD matter should account simultaneously for both types of cascade. This task seems arduous, essentially because the vacuum and medium-induced cascades evolve according to different evolution variables, the logarithm of the virtuality $t \equiv \ln Q^2\sim \ln \Tjet^2$ and the real time $t\sim L$, respectively. Moreover, we are facing a multi-scale problem: in addition to the hardest scale $Q^2$, which is typically of the order of $\sim$1000 GeV$^2$, the jet-medium interaction generates two additional global scales $Q_s^2 \ll Q^2$, which is of the order of $\sim 1 - 10$ GeV$^2$ and is related to transverse momentum broadening of the jet, and  $r_\perp^{-2}\equiv (\Tjet L)^{-2}$, where $r_\perp$ is the typical transverse size of the jet in the medium as depicted in  Fig.~\ref{fig:medium}. However, one might hope to achieve a consistent description of the in-medium parton showers in a certain limit owing to the fact that the vacuum and medium-induced showers are opposite in nature and well separated in angles: while the vacuum cascade is collimated and characterized by a suppression of large angle soft gluon radiation due to color coherence of the jet constituents, the medium-induced cascade forms at parametrically large angles where small angle hard gluon radiation is strongly suppressed due to another coherence effect the LPM effect that involve coherent scattering centers. Therefore, the situation might turn out to be less complicated than anticipated in certain cases. In the limit where $Q_s \ll r^{-1}_\perp$, which corresponds to the angular separation $\theta_c \gg \Tjet$, the medium does not resolve the inner structure of the jet and hence, is sensitive only to the total charge, that is of the parent parton. In this situation, the in medium and vacuum evolution can be factorized. The vacuum shower remains fully coherent and unmodified by the interaction with the medium apart from an overall energy lost at large angles \cite{CasalderreySolana:2012ef,Mehtar-Tani:2014yea}.  

\section{Interferences and decoherence}
\label{sec:decoherence}
\begin{figure}[htbp]\label{fig:antenna-vac}
\centering
\includegraphics[width=12cm]{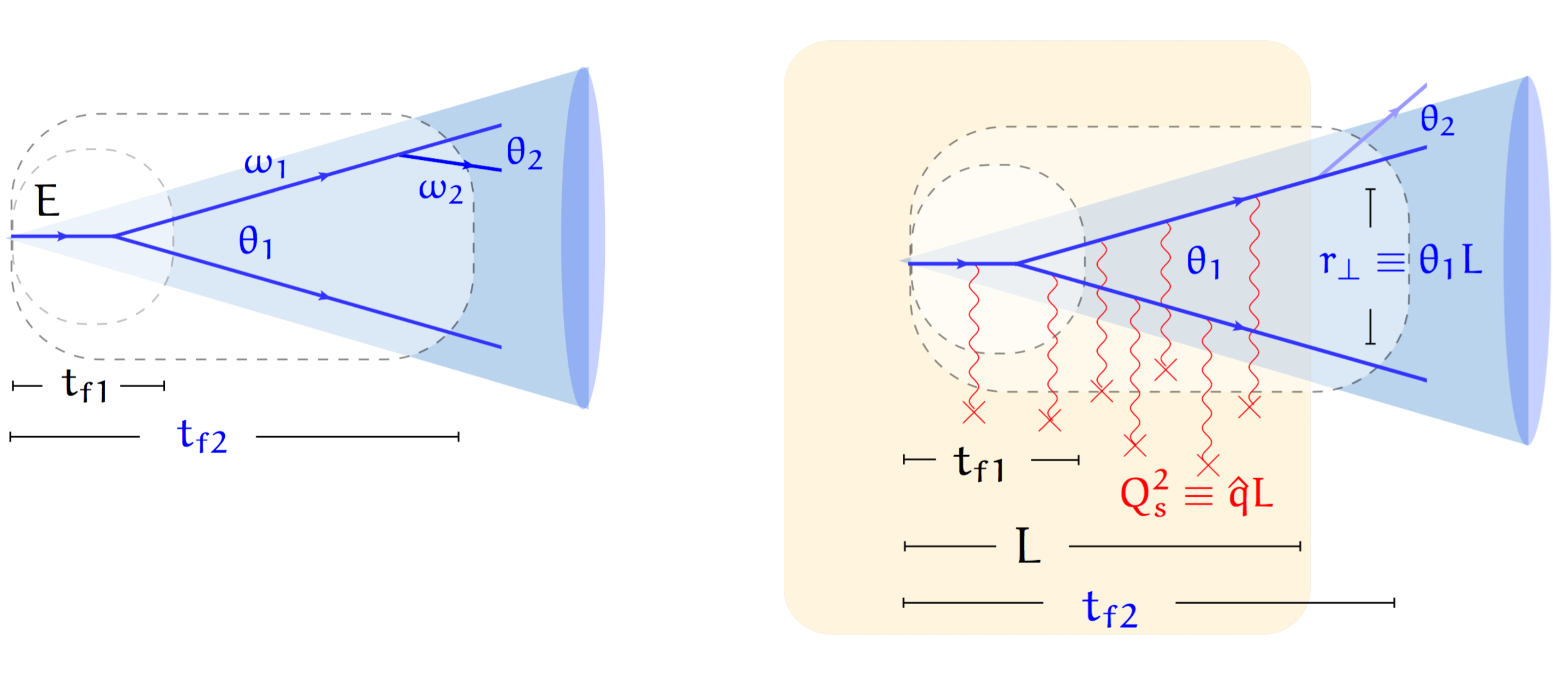}
\caption{Illustration of two successive branchings, strongly ordered in formation time: $t_{f1} \ll t_{f2}$. The parent gluon 0, with energy $E$,   radiates  gluon 1, with energy $\omega_1$, which further radiates a softer gluon 2 ($\omega_2$) (see text for details).  In-vacuum (left panel): destructive interferences between radiation off gluons 0 and 1 yields the angular ordering constraint that suppresses  large angle gluon radiation 2. In-medium (right panel): in the presence of a QCD medium, color randomization of gluon 0 and 1 (depicted by red vertical wavy lines) suppresses color coherence of the pair (antenna) and opens up phase-space for additional secondary radiation outside the medium.  }
\end{figure}

Generally, the separation of scales $Q_s \ll r_\perp^{-1}$ might not be sufficient and one might expect departures from the coherent jet picture. These corrections can then be computed perturbatively. In order to discuss corrections to the coherent picture, let us first briefly rediscuss the building block of jet evolution in vacuum: the antenna radiation pattern.  Consider two successive splittings that are strongly ordered in formation time $t_{f1} \ll t_{f2}$ and, for simplicity, also strongly ordered in energies $E\gg \omega_1 \gg \omega_2$. This process is illustrated in Fig.~\ref{fig:antenna-vac}, left panel. The probability for radiating gluon 1 is given by Eq.~(\ref{rad-prob}), $\rmd P_1\sim \bar\alpha\, \rmd \omega_1/\omega_1 \rmd \theta_1/ \theta_1$.  The radiation pattern of gluon 2, is modified however, due the interference between radiation off gluon 1 ($\omega_1$) and gluon 0 ($E$) that suppresses radiation at large angle (for inclusive observables the radiation off the total charge does not contribute due to real-virtual cancellations). After integration over the azimuth, the probability for the secondary splitting reads effectively 
\beq \label{rad-prob-2}
\rmd P_2\sim \bar\alpha  \frac{\rmd\omega_2 }{\omega_2}\frac{\rmd\theta_2 }{\theta_2} \Theta(\theta_1-\theta_2). 
\eeq
Note the ``strict'' angle-ordering constraint, $\theta_1>\theta_2$, that accounts for color coherence along parton shower.  

In the presence of a medium (cf. Fig.~\ref{fig:antenna-vac}, right panel), we assume that the first splitting that forms the antenna made of the gluons 0 and 1, occurs very early and the secondary radiation is much softer and thus occurs at much later times after the antenna has escaped the medium, namely, $t_{f1} \ll L \ll t_{f2}$. In this case the antenna suffers a rapid in-medium color precession that suppresses color coherence of the gluon pair. As a result, phase-space for large angle vacuum-like soft gluon radiation opens up \cite{MehtarTani:2010ma,MehtarTani:2011tz,MehtarTani:2011jw,CasalderreySolana:2011rz,MehtarTani:2011gf,MehtarTani:2012cy}, 
\beq\label{rad-prob-decoh}
\rmd P_2\sim \bar\alpha \frac{\rmd\omega_2 }{\omega_2}\frac{\rmd\theta_2 }{\theta_2}\left[ \Theta(\theta_1-\theta_2)+\Delta_\text{med}\Theta(\theta_2-\theta_1)\right], 
\eeq
with an intensity proportional to the decoherence parameter $\Delta_\text{med}=1-\exp[-r_\perp^2 Q_s^2/12]$, that interpolates between 0 in the fully coherent case, i,.e., $r_\perp \ll Q_s^{-1}$,  where the medium does not resolve the individual charges of the antenna, recovering Eq.~(\ref{rad-prob-2}), and 1 in the opposite case, i.e., $r_\perp \gg Q_s^{-1}$, which corresponds to full decoherence of the antenna system. The emergence of the antenna radiation pattern (\ref{rad-prob-decoh})  in a 1 to 3 gluon splitting was recently investigated \cite{Casalderrey-Solana:2015bww}. Note that the extra soft gluon radiation induced by color decoherence is vacuum-like, and does not get broadened by rescatterings since it occurs outside the medium. Therefore, it is of a fundamentally different nature than the large angle medium-induced soft gluon radiation discussed in section~\ref{sec:medium}. Such mechanism is expected to yield a softening of the jet inner structure. This provides a possible explanation for the excess of soft particle measured in the fragmentation function in Pb-Pb collisions at the LHC \cite{Aad:2014wha}, in addition to the recoil inside the jet cone of medium partons proposed recently \cite{He:2015pra}.  A generalization of the antenna radiation beyond the eikonal approximation was undertaken recently and is reported in these proceedings \cite{Apolinario:2015bfm}.  

\section{Higher-order corrections to jet-quenching observables }
\label{sec:NLO}
\subsection{1 to 3 splitting function and SCET} 
Going beyond the antenna setup to describe decoherence of the vacuum shower that develops inside the medium remains a challenge.  There have been recent attempts to compute the full 1 to 3 splitting function in the single \cite{Fickinger:2013xwa} and multiple scattering approximations \cite{Arnold:2015qya}. In the latter, only a subset of diagrams was considered and the authors where able to recover the double logarithmic radiative correction to energy loss that we shall discuss at length in the next section. The 1 to 3 gluon splitting was also analyzed recently assuming a strong ordering of the two successive 1 to 2 splittings that form the 1 to 3 splitting, in energies and formation times \cite{Casalderrey-Solana:2015bww}. This work supports the description of the collimated jet shower in the presence of a medium as a collection of antennas.

Furthermore, there is a growing interest in using Soft-Collinear-Effective-Theory \cite{Bauer:2000yr}, to deal with the increasing complexity of calculations beyond leading-order. It has proven to be useful in resuming Next-to-Leading-Log corrections to Jet shapes for instance \cite{Chien:2014nsa}. 

\subsection{Radiative corrections to momentum broadening} 

Recently, it was shown that radiative corrections to transverse momentum broadening exhibit a double-logarithmic (DL) enhancement  of the length of the medium \cite{Wu:2011kc,Liou:2013qya,Blaizot:2013vha},
\beq
\langle k_\perp^2\rangle_\text{typ} \simeq \hat q L \left[1+\frac{\bar\alpha}{2}\ln^2\frac{L}{\tau_0}\right], 
\eeq 
where the cut-off $\tau_0\sim 1/m_D$ is of the order of the in-medium screening length. This double logarithm arises from the the integration over the life time, $\tau$,  of the gluon fluctuation from $\tau_0$ to $L$ and over its transverse momentum in the regime of a single scattering $k_\perp \gg  \hat q \tau $. The region $k_\perp \lesssim \hat q \tau $ corresponding to the LPM suppression regime where multiple scatterings are important, do not contribute to the double logarithmic enhancement. 

These potentially large non-local corrections might question the validity of the probabilistic picture discussed above. We shall see in the next section that these radiative corrections can be absorbed in a redefinition of the jet-quenching parameter and the probabilistic picture remains valid.   

Finally, in a slightly different context, namely,  Deep-Inelastic Scattering off a heavy nucleus, NLO corrections to transverse momentum broadening of the final state quark were computed in the Higher-Twist framework \cite{Kang:2013raa,Kang:2014ela}. In this work, no double logarithmic enhancement was reported. 

\section{Renormalization of the jet-quenching parameter }
\label{sec:RG-qhat}

Although the large DL radiative corrections discussed previously are non-local, the dominant contribution (owing to the logarithmic form of the corrections) is effectively given by independent multiple radiative corrections associated with individual scatterings in the multiple scattering approximation.  This allows us to absorb the double logarithmically enhanced radiative corrections into a renormalization of the transport coefficient, $\hat q $ \cite{Blaizot:2014bha,Iancu:2014kga}. It follows that radiative corrections to the mean radiative energy loss yields the same DL enhancement \cite{Blaizot:2014bha,Wu:2014nca}, 
 \beq
\langle \Delta E\rangle  \sim  \hat q L^2 \left[1+\frac{\bar\alpha}{2}\ln^2\frac{L}{\tau_0}\right], 
\eeq 
demonstrating the universal nature of the double-logarithmic radiative corrections. 

For large media, when $\bar\alpha \ln^2(L/\tau_0)\sim1$, one has to resum the DL power corrections. The resummation of the leading logarithms involves large separation of scales between successive gluonic fluctuations which are ordered in formation time, $\tau_0 \ll \tau _1\ll ...\ll\tau_n\equiv \tau_{\text{max}}\sim L $ and in transverse momentum $m_D  \ll \q _1\ll ...\ll\q_n\equiv \k $. The essential difference with the standard Double-Logarithmic Approximation (DLA) lies in the reduced DL phase-space set by the LPM effect, i.e., multiple-scatterings: in the DLA only a single scattering contributes, which imposes  that the life time of a fluctuation of frequency $\omega$ must be smaller than the BDMPS formation time: $\tau < \sqrt{\omega/\hat q} $. In terms of our variables ($\tau$ and $k_\perp$), $k_\perp \gg  \hat q\,\tau $. The resummation of the DL's can be formulated in terms  of the evolution equation, 
\beq\label{qhat-evol}
\frac{\del \hat q (\tau,\k^2)}{\del \log \tau} = \, \int_{\hat q  \tau }^{\k^2}\frac{d\q^2}{\q^2}\, \bar \alpha(\q)\, \hat q (\tau,\q^2).
\eeq
This equation describes the evolution of the jet-quenching parameter from an initial condition $\hat q_0\equiv\hat q (\tau_0)$, which can be computed on the lattice \cite{Majumder:2012sh,Panero:2013pla}, or in the  Hard-Thermal-Loop framework where recently $\hat q$ was calculated to Next-to-Leading-Order \cite{Ghiglieri:2015ala}. Note the running of the coupling that explicitly encodes the fact that the coupling increases with decreasing transverse momentum from a hard scale $Q_s\sim \hat q L$ down to $m_D$. Remarkably, due to the QCD evolution of the jet-quenching parameter, the mean radiative energy loss exhibits an anomalous scaling. In the limit of large media, we obtain,   
\beq 
\langle \Delta E\rangle \sim L^{2+\gamma}\, \quad \text{with}\quad \gamma=2\sqrt{\bar\alpha}\,,
\eeq  
which lies between the standard small coupling result, $\langle\Delta E \rangle \sim L^2 $ and the strong coupling result obtained in the framework of the AdS/CFT correspondence, $\langle \Delta E \rangle \sim L^3 $ \cite{Hatta:2007cs}.

\section*{Acknowledgement}
We would like to thank N. Armesto, J.~P. Blaizot, J. Casalderrey-Solana, F. Dominguez, L. Fister, E. Iancu, M. Martinez, J. G. Milhano, A. H. Mueller, C. A. Salgado, K. Tywoniuk, M. A. C. Torres and B. Wu for many fruitful discussions and collaborations. This work was supported by the U.S. Department of Energy under Contract No. DE-FG02-00ER41132. 




\nocite{*}
\bibliographystyle{elsarticle-num}
\bibliography{martin}



\end{document}